# Nonmonotonic photostability of BA$_2$MA$_{n-1}$Pb$_n$I$_{3n+1}$ homologous layered perovskites

*Natalia N. Udalova [1], Sergey A. Fateev [1], Elizaveta M. Nemygina [1], Andrea Zanetta [2], Giulia Grancini [2], Eugene A. Goodilin [1], Alexey B. Tarasov [1,*]*

[1] Laboratory of New Materials for Solar Energetics, Department of Materials Science, Lomonosov Moscow State University, Lenin Hills, 119991, Moscow, Russia.

[2] Department of Chemistry and INSTM, University of Pavia, via Taramelli 16, 27100 Pavia, Italy.

ABSTRACT

Layered lead halide Á$_2$A$_{n-1}$Pb$_n$I$_{3n+1}$ perovskites (2D LHPs) are attracting considerable attention as a more stable alternative with respect to APbI$_3$ parent phases – a promising material for a new generation of solar cells. However, a critical analysis on the photostability of 2D perovskites homologous row is still lacking. In this work, we perform a comparative study of BA$_2$MA$_{n-1}$Pb$_n$I$_{3n+1}$ (BA – butylammonium, MA – methylammonium) 2D LHPs with different layer number n = 1-3, considered as study-case systems, and MAPbI$_3$, as a reference. We discuss a stability testing protocol with general validity, comparing photometrical determination of iodine-containing products in nonpolar solvents, X-ray diffraction, and photoluminescence spectroscopy. The resulting photostability trends in inert atmosphere demonstrate a nonmonotonic dependence of degradation rate on the perovskite layer thickness n with a "stability island" at n ≥ 3 which is caused by a combination of antibate factors of electronic structure and chemical composition in the family of 2D perovskites. We also identify a critical oxygen concentration in the surrounding environment that affects the mechanism and strongly enhances the rate of layered perovskites photodegradation.

INTRODUCTION

Hybrid lead halide perovskites APbX$_3$ (A – organic cation, e.g. methylammonium = MA$^+$, X – halide anion I$^-$, Br$^-$, Cl$^-$) are world recognized promising materials for advanced optoelectronics including solar cells[1], light-emitting diodes[2], transistors[3], X-ray detectors[4]. This class of materials (referred as 3D perovskites) possesses impressive optical and electrical properties[5] along with being fabricated with simple and low-cost preparation methods.[6] However, one of the main obstacles prior mass commercialization is their low stability with respect to natural factors such as heat stress, light-soaking, high humidity conditions, poor resistance to oxygen exposition, electrical bias cycling[7–10]. Layered lead halide perovskites (2D LHPs) have recently emerged as a promising alternative to overcome stability problems of 3D parent phases.[11,12] The 2D LHPs with a general formula $Á_{2/q}A_{n-1}Pb_nX_{3n+1}$ are layered structures consisting of $[A_{n-1}Pb_nX_{3n+1}]^{2-}$ perovskite-like layers (structural blocks) with singly (q = 1) or doubly (q = 2) charged $Á^{q+}$ organic cations in the interlayer space.[13] The number $n$ corresponds to the number of the inorganic layers of corner-sharing octahedra $[PbX_6]$ held together by the small $A^+$ cation if n > 1.

The high stability of 2D LHPs against moisture due to the presence of hydrophobic Á cations is consistently demonstrated when they are integrated in perovskite solar cells (PSCs).[14–16] However, the reported results regarding the photostability of layered perovskites still remain limited and controversial (see Table S1 in ESI). Along with the data on superior stability of PSCs based on 2D LHPs with n = 4 or 5, as compared to 3D-based devices[17,18], worst photostability of $PEA_2PbI_4$ single crystal flakes with respect to monocrystalline $MAPbI_3$ is noted.[19] Such misleading results demand a systematic study and much deeper understanding of the degradation mechanisms and issues affecting the photostability of layered perovskites with different layer number $n$. For the layered perovskites with $n$ = 3-5, the most demanded for devices, only a few works have provided a comprehensive stability study (ambient, thermal and photostability) for different 2D LHPs, however, without addressing the dependence of the stability on the layer number $n$.[20] Additionally, there is scarce information on the effect of the surrounding atmosphere (inert or oxidizing) during light soaking on the photostability of 2D LHPs, being, on the contrary, a well-studied topic for the 3D parent phases.[21,22] To tackle this challenge, an in-depth stability analysis and novel testing protocols should be developed for 2D LHPs.

In this article, we provide a detailed photostability study of layered perovskites $BA_2MA_{n-1}Pb_nI_{3n+1}$ (BA = butylammonium) with n = 1 – 3 and compare the results with 3D parent phase $MAPbI_3$. First, we implement a combination of various testing protocols and discuss the advantages and limitations of different stability characterization methods such as photometrical determination of iodine-containing products in nonpolar solvents, X-ray diffraction, and photoluminescence spectroscopy. Second, we carefully investigate the influence of the amount of oxygen in the surrounding atmosphere during aging comparing 2D and 3D perovskites degradation mechanism and its rate. Such a joint stability study of layered perovskites with various n and 3D perovskite allows to develop optimal photostability assessment protocol equally applicable for two classes of LHPs and "build a bridge" in understanding of their degradation mechanisms. Finally, we compare the photostability of the individual perovskite materials within solar cells to assess the relevance of our findings in operational devices.

RESULTS AND DISCUSSION

The $BA_2MA_{n-1}Pb_nI_{3n+1}$ layered perovskites are among the most known 2D perovskite used for stable solar cells in phase-pure or 2D/3D mixed configurations.[23,24] Hereinafter, we refer $BA_2PbI_4$ as n1, $BA_2MAPb_2I_7$ as n2, $BA_2MA_2Pb_3I_{10}$ as n3, and $MAPbI_3$ as MAPI. Thin films were prepared by a simple solution-based approach via spin-coating (details are provided in the experimental section in ESI). All thin film samples were initially characterized by a combination of XRD, photoluminescence spectroscopy (PL), UV-vis absorption spectroscopy, and energy-dispersive X-ray spectroscopy (EDX) as shown in Figure S1 - S3 and Table S2 (ESI). According to XRD, adsorption and PL spectroscopy, almost all the perovskite films are single phase except n2 and n3 having small n1 and n2 phase impurities, respectively, as shown by minor additional

XRD reflexes (Figure S1) and PL peaks (Figure S2). Further we provided a complex stability assessment of perovskite samples by several testing protocols and under various atmospheres, listed in Table 1.

**Table 1**. Conditions of all stability testing experiments used in the current work.

| | Type of environment | | | |
|---|---|---|---|---|
| | nonpolar solvent (decane / degassed decane) | dry air (~ $2*10^5$ ppm $O_2$) | 99.9% Ar (~100 ppm $O_2$) | 99.9996% Ar (< 10 ppm $O_2$) |
| **Stability test type** | Light-soaking (blue LED, 450 nm, 100 mW/cm$^2$) | | | Light-soaking (Blue LED, 450 nm, 100 mW/cm$^2$) |
| | | | | Thermal heating (100°C) |
| **Methods used** | UV-vis abs. of decane; XRD; SEM | XRD | XRD | XRD; PL |

**Perovskite photostability measurement in nonpolar solvent**

One of the most developed method for the analysis of LHPs photodegradation relies on in-situ photometrical determination of iodine-containing products in nonpolar solvents.[25,26] Here we apply for the first time a similar approach for layered perovskites with different n. The scheme of the experimental setup is given in the inset of **Figure 1b**. Based on the recently reported classification of solvents for hybrid perovskites,[27] decane was selected as the most appropriate nonpolar solvent for this set of experiments. Also, it forms no charge-transfer complexes with molecular iodine (in contrast to toluene, benzene, etc.) and should not enhance iodine extraction during the stability test. At first, we confirmed that decane does not show any dissolution or interaction with n1 and MAPI perovskites for 24 h upon 70°C heating in darkness (Figure 1a). On the other side, after only 50 min of light-soaking (blue LED with 450 nm and 100 mW/cm$^2$) we observe a distinct release of molecular iodine, manifested by the absorption line at 520 nm in both samples. In particular, the absorption intensity of $I_2$, released from the 2D perovskite sample, is found ~20 times higher than that from the 3D perovskite. Additionally, in the case of n1 perovskite, we observe a second absorption peak around 360 nm which could be referred to polyiodide species,[28,29] previously detected during 3D perovskites degradation.[30,31] The same peak with a much lower intensity is observed for n2 sample while for n3 and MAPI it is absent (Figure S4). The graphical illustration of iodine release dynamics during light-soaking of perovskites with different *n* number is provided in Figure 1b (raw data is given in Figure S4-S5, ESI). To minimize the impact of film thickness we calculated the exact fractions of released iodine, normalized on the amount of illuminated perovskite (for details see the Experimental Section). The resulting graphs are shown in Figure 1c.

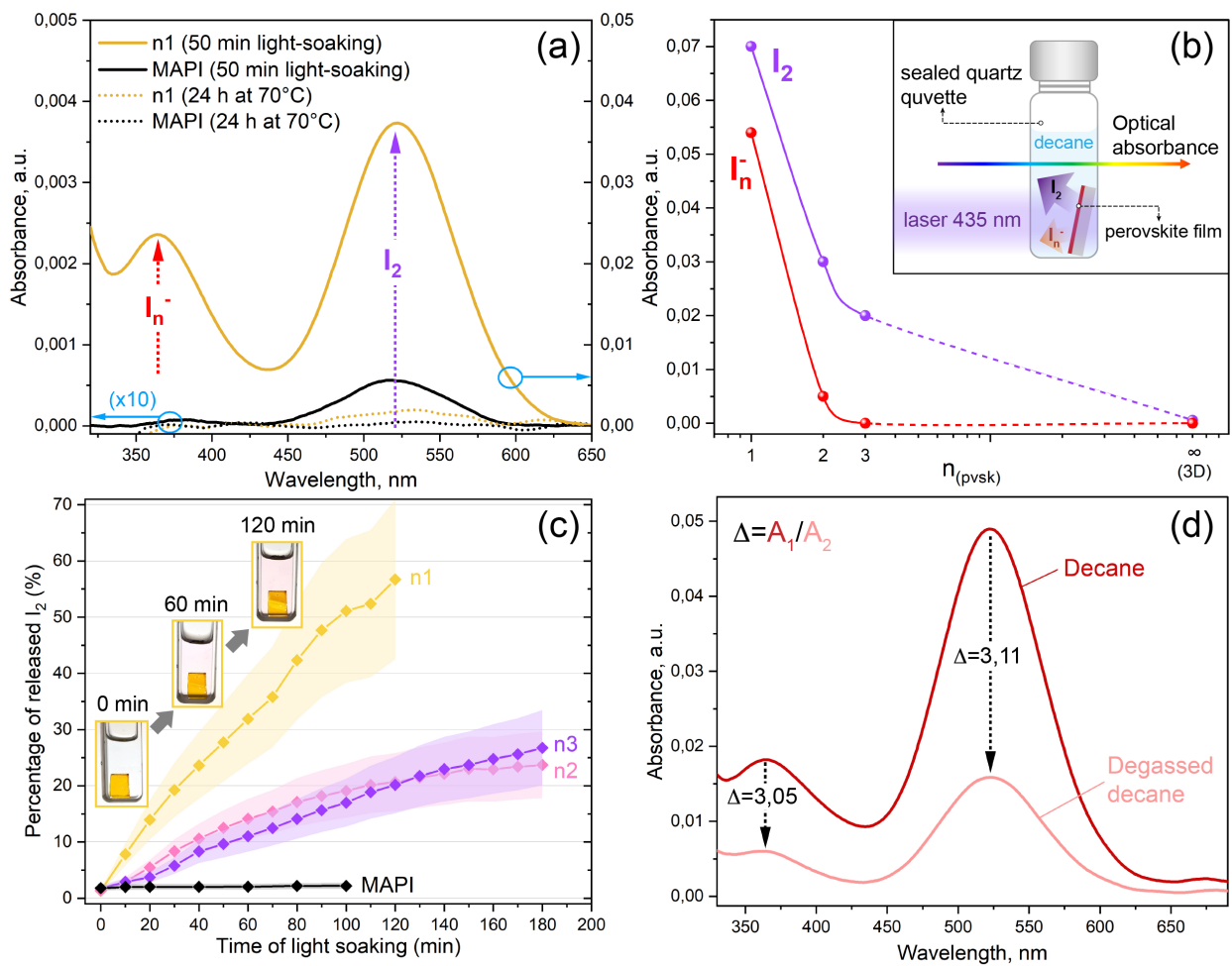

**Figure 1.** (a) Absorption spectra of decane solutions above MAPbI$_3$ (black lines) and BA$_2$PbI$_4$ (yellow lines) films after 24 h of heating at 70°C (dashed lines) and 50 min of light-soaking with blue light (solid lines). (b) Absorption intensities of I$_2$ and I$_n^-$ degradation products for BA$_2$MA$_{n-1}$Pb$_n$I$_{3n+1}$ perovskite with different *n* after 100 min of light-soaking in decane. The inset in (b) shows a schematic representation of in-situ setup for photometrical determination of perovskite decomposition products released into decane upon light-soaking. (c) The resulting time dependencies of the released I$_2$ fraction in % (normalized on the amount of irradiated perovskite) during light-soaking (435 nm, 100 mW/cm$^2$) of hybrid perovskites n1, n2, n3, and MAPI in decane. The inset photographs demonstrate the decane color change at different irradiation time of n1. (d) Absorption spectra of decane above n1 film after 60 min of light-soaking in pristine decane (dark red) and decane degassed in a vacuum chamber (light red). The parameter $\Delta = A_1/A_2$ equals to the ratio of corresponding absorption peak intensities for pristine (A$_1$) and degassed decane (A$_2$).

We can rationalize this finding as that the molecular iodine release rate in decane is inversely dependent on the thickness of perovskite slabs *n* thus evidencing for the following photostability sequence: MAPI > n3 ~ n2 > n1. This tendency is also supported by SEM data of the degraded samples, as shown in Figure S6, since the morphology of the least degraded sample corresponds to MAPI and the most degraded – to the n1 sample. Importantly, XRD shows that light-soaking of hybrid perovskites in decane gives no PbI$_2$ phase (Figure S7). Instead, a noticeable intensity drop of diffraction peaks without their widening is observed for the degraded sample which might correspond to the reduction of the amount of the initial perovskite phase. Further in the text, we will use a semi-quantitative measure of perovskite phase degradation as a ratio of current XRD peak area A (after stability test) to the initial value $A_0$ (before the stability test) designated as "$A/A_0$". The smaller this parameter, the more intense degradation is supposed to happen.

The distinctive feature of decane spectra above the n1 sample is a noticeable absorption peak at ~360 nm that could be explained by the non-equilibrium photochemical release of polyiodide species consisting of $BA^+$ cations together with $I^-$ counter ions initially forming the metastable solution and then resulting in a yellowish sediment for 24 hours of storage (Figure S8). This organic cation extraction into a nonpolar solvent is not observed for a more polar $MA^+$ but is possible for sufficiently hydrophobic cations like $BA^+$. With an increase of 2D perovskite *n* number, the proportion of $MA^+$ in perovskite composition rises decreasing the possibility of BAI extraction. Notably, the polyiodide absorption steeply decreases moving from n1 to n2, and for the n3 sample (MA/BA = 1) no $I_n^-$ absorption is already detected as well as for the MAPI phase (Figure 2b, Figure S4). Based on these observations, we assume that BAI extraction in decane may enhance degradation of layered perovskites immersed in a nonpolar solvent upon light-soaking. This effect is more pronounced for the studied layered perovskites with lower *n* and also for 2D perovskites based on hydrophobic cations; therefore, in situ photometry of a nonpolar solvent seems to be insufficiently universal and should be applied with cautions for 2D perovskites.

In addition, we revealed an influence of dissolved oxygen in decane on the rate of light-induced perovskites degradation. The amount of $O_2$ in 2 ml of decane under standard conditions (25°C, 100 kPa) is about $2*10^{-5}$ mol[32] which is two orders of magnitude higher than the estimated amount of perovskite in the immersed films used in this experiment ($2-3*10^{-7}$ mol), leading to the involvement of $O_2$ into perovskite photodegradation rate.[33] We found that degassing of decane before the light-soaking experiment allows to slow down the release of both iodine-containing products ($I_2$ and polyiodide) by three times (Figure 1d). Moreover, we revealed that the mechanism of perovskite light-induced degradation in non-polar solvent differs from the one under the ambient atmosphere. According to absorption spectroscopy of n2 perovskite films before and after light-soaking in different environment, the material undergoes ~20 times faster photodegradation in ambient air than in decane (Figure S9, ESI). In particular, already after 10 min of light soaking in air, we observe a noticeable growth of absorption edge at 2.36 eV corresponding to $PbI_2$ phase which is not the case for the sample irradiated in decane.

Summarizing all the data from hybrid perovskites stability tests in nonpolar solvents, we should note that despite the reliability of this approach for studying the degradation of 3D perovskites with small polar organic cations ($MA^+$, $FA^+$), it is hardly suitable for layered perovskites containing hydrophobic cations. One more argument against this approach is a difference in perovskite degradation mechanism in nonpolar solvents comparing to the one in a gaseous environment. In this context, to properly assess the photostability of hybrid perovskites with different dimensionality, we suggest to perform a periodic tracking of the perovskite photoluminescence signal during light or heat aging. An essential step preceding these experiments is the definition of the requirements of the surrounding atmosphere.

**Perovskite photostability measurements in the atmosphere with different oxygen content**

Our results demonstrate that light-soaking effects in dry air (~$2*10^5$ ppm of $O_2$) and in an inert glove box (<10 ppm of $O_2$) differ drastically. In the case of inert atmosphere, all sets of samples retain their initial color for at least 16 h of the constant blue light irradiation with power density 100 mW/cm$^2$ (Figure S10). In the case of dry air, we observe a complete degradation of all 2D perovskites regardless of *n* leading to light-yellow color already after 20 min under the same light-soaking conditions, while MAPI films retain at most initial color with only local yellowing, highlighted in Figure S10 with red circles. To reveal the influence of oxygen concentration on the photodegradation mechanism and to identify a critical $O_2$ concentration, we analyzed phase composition and crystal structure of n1 and MAPI perovskite films by XRD after aging under three types of atmospheres: (1) dry air with ~$2*10^5$ ppm $O_2$, (2) argon with 99.99% purity (~100 ppm $O_2$), and (3) argon-filled glove box with < 10 ppm $O_2$. The duration of light-soaking was increased from 5 h for (1), to 83 h for (2), and 150 h for (3) in order to diminish the impact of different degradation rate of hybrid perovskites, known to be dependent on the oxygen amount in the case of 3D perovskites.[34] The XRD patterns of n1 and MAPI samples are provided in **Figure 2a** and 2b, respectively. In the case of maximal oxygen concentration in the atmosphere ($2*10^5$ ppm), we observe a pronounced difference in behavior of 2D and 3D perovskites: after 5 h of light-soaking the n1 sample undergoes almost complete phase degradation with the decrease of incident perovskite (200) diffraction peak area by 4 orders of magnitude ($A/A_0=7*10^{-4}$) and also a widened $PbI_2$ peak at 12.6° appears (Figure 2a). In contrast, the MAPI sample reveals only a ~3 times decrease of diffraction intensity and demonstrates negligible signs of the $PbI_2$ phase ($PbI_2$/pvsk = 0.14) (Figure 2b). We suppose that a more rapid photodegradation of the 2D perovskite under oxidizing environment is caused by a much easier penetration of oxygen molecules into the layered structure as compared to a denser, interlinked framework of the 3D perovskite structure (Figure 2c and d). This result is consistent with the UV-light stability study in air of perovskite solar cells based on MAPI and n3 samples in which 2D-PSCs demonstrate a pronounced decrease of PCE for only 30 min of light-soaking due to a strong $V_{oc}$ decrease while the MAPI-based device shows no sign of degradation (Figure S11).

A decrease of the $O_2$ amount down to 100 ppm almost completely diminishes the discrepancy between 2D and 3D perovskites on the time scale of 83 h of the photostability test. In both cases, we observe a complete disappearance of diffraction reflexes from perovskite phases and formation of $PbI_2$ as also seen by the samples color change (Figure S12). Interestingly, the widening of the $PbI_2$ (001) reflex monotonically increases with a decrease of the *n* number (Figure S13). We suppose that this behavior might be caused by the formation of disordered $PbI_2$ ultrathin flakes during the degradation of the layered structures with a low thickness of perovskite slabs (n= 1-3).

With a further decrease of oxygen content down to < 10 ppm, only a trace amount of $PbI_2$ phase is detected in the n1 sample ($PbI_2$/pvsk = 0.33) while no sign of $PbI_2$ formation can be found at all in the MAPI film after 150 h of light-soaking. The initial perovskite diffraction peaks are preserved in both samples with a decrease of the absolute intensity by 3 orders of magnitude for n1 and by only ~3 times for MAPI (Figure 2a, b). Performing UV-light soaking of PSCs in an inert glove box demonstrates that, in contrast to the ambient atmosphere, both types of PSCs undergo a monotonous $J_{sc}$ and $V_{oc}$ decrease (Figure S14). However, PCE measurements of unencapsulated devices in this series were made outside the box which could interfere the degradation mechanism of hybrid perovskite layers. Indeed, the most pronounced PCE decrease is observed in the n3 based PSCs comparing to the MAPI device.

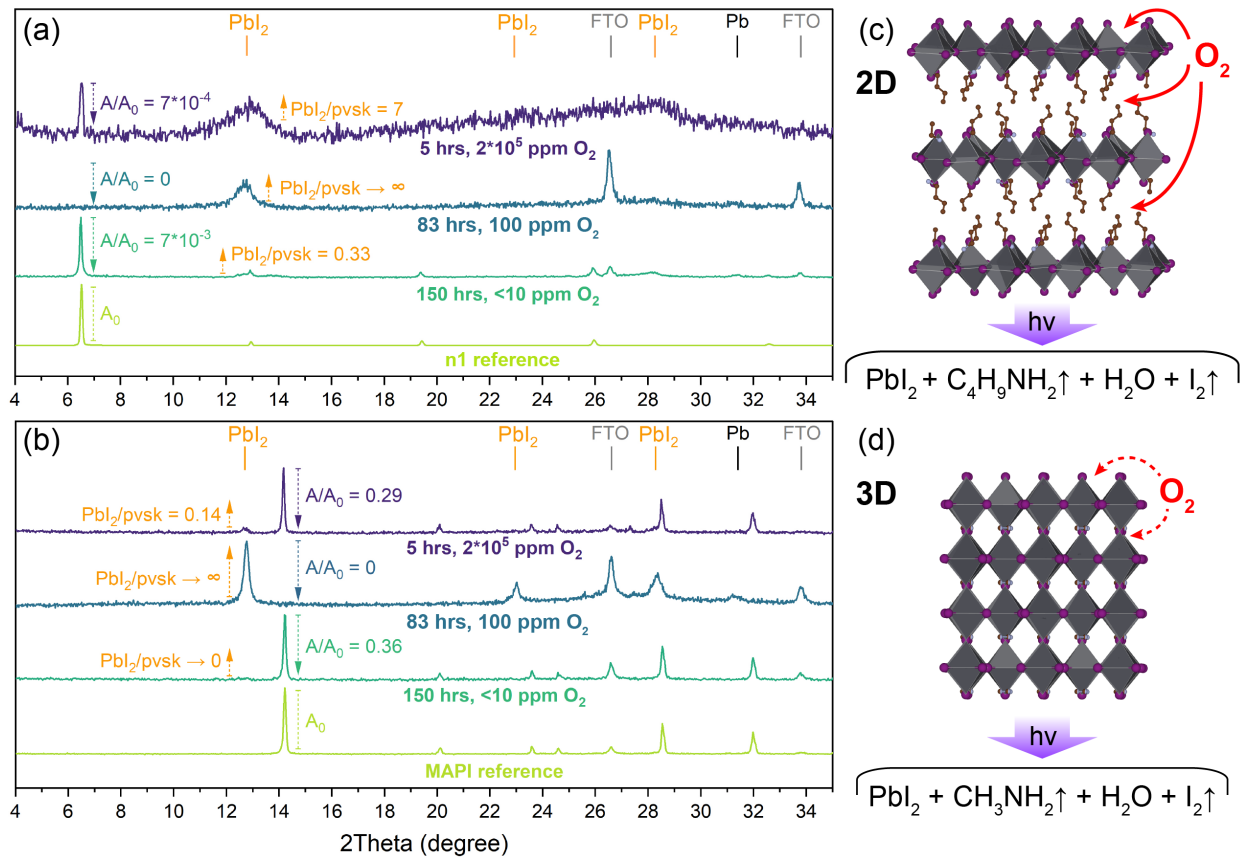

**Figure 2.** (a, b) Normalized X-ray diffraction patterns of n1 (a) and MAPI (b) samples before and after light-soaking under 100 mW/cm² white or blue LED in atmosphere with different $O_2$ concentration (< 10 ppm, 100 ppm, and $2*10^5$ ppm). The decrease of perovskite diffraction peak area with respect to the reference sample is denoted as "$A/A_0$". The ratio of $PbI_2$ (100) peak area at 12.6° to perovskite peak area ((200) at 6.5° for n1; (110) at 14.2° for MAPI) in each sample is shown as "$PbI_2$/pvsk". (c, d) Schematic illustration of oxygen diffusion paths into 2D (c) and 3D (d) perovskite crystal lattice with listed degradation products for both cases.

As a result, we found that oxygen in high or even moderate concentrations strongly affects the photodegradation mechanism of layered perovskites, increasing their corrosion rate by an order of magnitude as compared to MAPI. With lowering the $O_2$ concentration down to 100 ppm (99.99% Ar), the ongoing mechanism is not affected since the main degradation product remains $PbI_2$ in both atmospheres. On the contrary, the decrease of $O_2$ down to < 10 ppm leads to a drastic change in perovskite degradation mechanism – we observe a negligible color change of samples and almost no signal from $PbI_2$ regardless of the perovskite *n* number. We consider that XRD and even optical absorption are not sensitive enough for a suitable study of perovskite slow photodegradation processes under the purely inert environment. At the same time, this type of atmosphere is the most relevant for stability tests because perovskite solar cells are usually assembled and encapsulated in inert glove boxes with low $O_2$ and $H_2O$ concentration. Therefore,

PL spectroscopy is expected to be a much more sensitive and informative method to assess stability of various hybrid perovskites.

**Perovskite thermal and photostability tracking in inert atmosphere using PL spectroscopy**

We simultaneously perform two types of stability tests – light-soaking under blue LED (450 nm, 100 mW/cm$^2$) and thermal heating at 100°C (**Figure 3a**), in order to distinguish light and heat-induced processes in hybrid perovskites. As an analytical signal we use an average intensity of perovskite PL peak, collected from multiple dots, normalized to the initial average PL intensity of each sample at the beginning of the stability experiment. The results of both tests are shown in Figure 3b for light-soaking and 3c for thermal heating. Original PL spectra after averaging are given in Figure S15 (ESI). According to the results of light-soaking, we observe the following photostability trend: n3 > n2 ~ MAPI > n1. Notably, this result does not agree with the previously observed stability trend in decane: MAPI > n3 ~ n2 > n1. The strongest controversy is found for 3D perovskite MAPI appeared to be the most stable in decane, but being one of the least stable material in argon atmosphere (Figure 3b). At the same time, the layered perovskites follow the overall trend of increasing photostability with the growth of the *n* number (Figure 1c and 3b). In the case of heat-induced degradation, we observe a less pronounced difference in relative distribution of PL intensity with the following stability trend: n1 > n3 > n2 > MAPI. Both photo- and thermal stability trends are proved to be reproducible on the basis of two sequentially conducted stability tests in the same conditions (Figure S16, ESI).

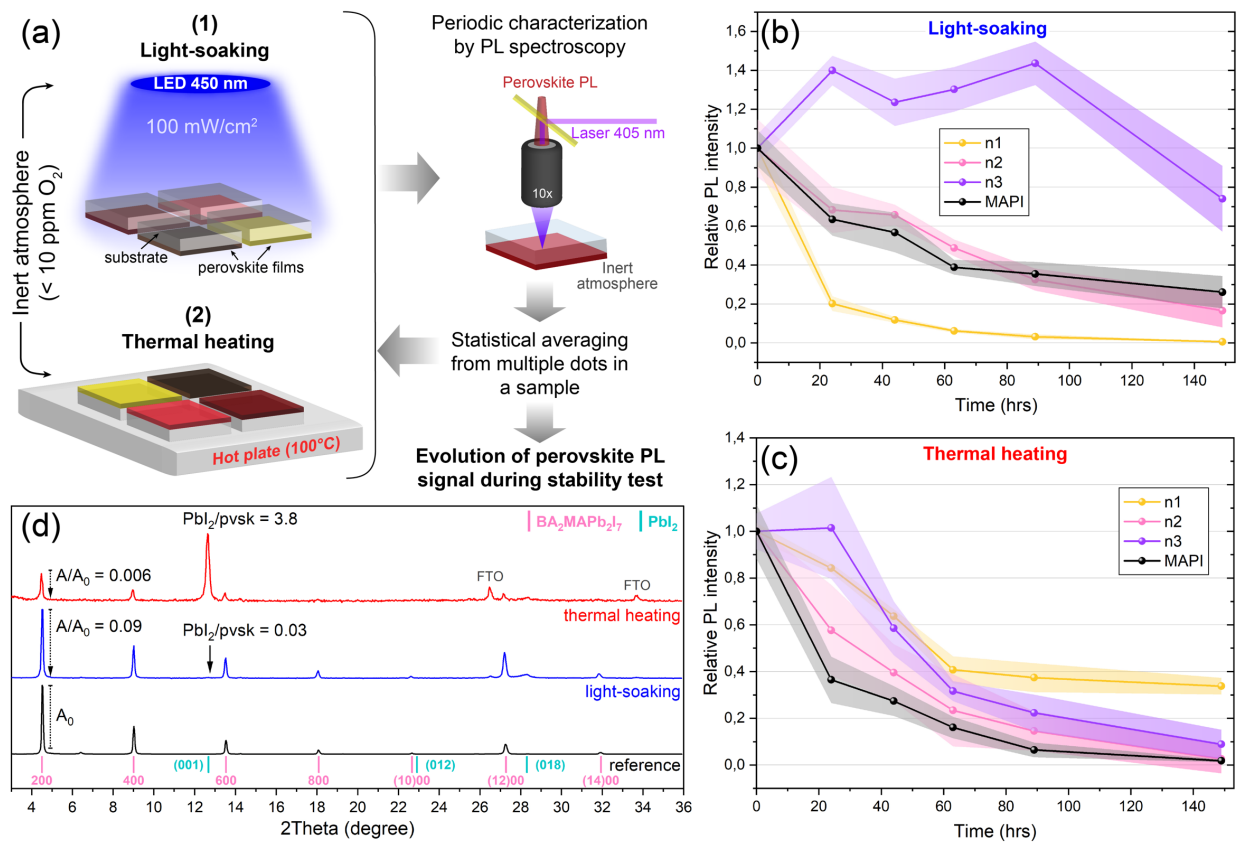

Figure 3. (a) The scheme of stability light-soaking (1) and thermal heating (2) tests in an inert atmosphere with periodic characterization of perovskites using the PL spectroscopy. (b, c) The resulting time evolution

graphs of the PL intensity for a series of perovskite samples, tested under blue LED light (b) and thermal heating at 100°C (c). (d) Diffraction patterns of pristine and degraded n2 sample. Colored vertical sticks with (hkl) indexes correspond to $BA_2MAPb_2I_7$ (pink) and $PbI_2$ (cyan) phases. "$A/A_0$" is a coefficient of perovskite (200) peak area decrease with respect to the reference n2 sample. "$PbI_2$/pvsk" – ratio of $PbI_2$ diffraction peak area to the perovskite (200) peak in each sample.

The resulting XRD patterns for the n2 sample (Figure 3d) confirm the principal difference of hybrid perovskites degradation mechanisms under heating and under light irradiation: thermal degradation is mainly caused by the loss of volatile species ($CH_3NH_2$, HI, $C_4H_9NH_2$) with crystallization of the $PbI_2$ solid phase while photodegradation seems to be driven by the combination of photoelectrochemical reactions, discussed in more details below. The same features in X-ray diffraction patterns are observed for all other samples (Figure S17, ESI). A slightly higher resistance of layered perovskites to heating can be explained by much lower volatility of butylamine (vapor pressure at 20°C $P_{vap}$ = 9.1 kPa) compared to methylamine ($P_{vap}$ = 186.1 kPa).

**A possible photodegradation mechanism**

At first, to rationalize the experimentally observed nonmonotonic dependency of perovskites photostability in inert atmosphere according to the row n3 > n2 ~ MAPI > n1, a generalized light-induced degradation mechanism should be discussed. During light absorption, charge carriers in hybrid perovskites are being excited and involved in a combination of possible photochemical redox reactions such as: (1.1) oxidation of iodide ions to molecular iodine; (1.2) reduction of $Pb^{2+}$ to metallic lead; (1.3) reduction of alkylammonium cations to amine and hydrogen molecules (Table 2). In addition to these processes several non-redox reactions are possible: (1.4) formation of polyiodide and (1.5) deprotonation of organic cation by iodide with the release of HI and amine. In accordance with current experimental results and some literature data we can assume that reactions (1.1) and (1.2) as well as (1.5) are the main degradation driving processes while reduction of organic cation (1.3) is unlikely due to the limitations in electron transfer from inorganic [$PbI_3$] sublattice to weakly bound $RNH_3^+$ cation. Polyiodide formation was observed earlier for 3D perovskites[30,31] and in this work for n1 and n2 layered perovskites during light-soaking in decane (Figure 1a-b). However, this process has a secondary order and depends on iodine vapor pressure above perovskite under light with accompanying heating.

As a result, we can assume that intensity of ongoing redox reactions strongly depends on the energy of charge carriers in the material, which increases along with bandgap. Indeed, in the case of 2D LHPs we observe increasing photostability in the row n1 ($E_g$=2.4 eV) < n2 (2.1 eV) < n3 (1.98 eV); however, 3D perovskite MAPI with the lowest $E_g$ (1.6 eV) demonstrates photostability close to n2 sample. We explain this stability decrease by a higher volatility of $MA^+$, fraction of which grows with n number and, at some point, this factor will dominate.

**Table 2**. Possible half-reactions ongoing in LHPs during light-induced degradation in the case of inert and oxidizing environment.

| № | Half-reaction in inert atmosphere | № | Half-reaction in oxidizing atmosphere |
|---|---|---|---|
| 1.1 | $2I^- + 2h^+ \rightarrow I_2^{(g)}$ | 2.1 | $2I^- + 1/2O_2 \rightarrow I_2^{(g)} + O^{2-}$ |
| 1.2 | $Pb^{2+} + 2e^- \rightarrow Pb^0$ | 2.2 | $O_2^{(g)} + 4e^- + 4RNH_3^+ \rightarrow 2H_2O + 4RNH_2^{(g)}$ |
| 1.3 | $RNH_3^+ + e^- \rightarrow RNH_2^{(g)} + 1/2H_2^{(g)}$ | 2.3 | $Pb^{2+} + 2I^- \rightarrow PbI_2^{(s)}$ |
| 1.4 | $I_2 + I^- \rightarrow I_3^-$ | 2.4 | $Pb^{2+} + O^{2-} \rightarrow PbO^{(s)}$ |
| 1.5 | $RNH_3^+ + I^- \rightarrow RNH_2^{(g)} + HI^{(g)}$ | 2.5 | $O^{2-} + 2RNH_3^+ \rightarrow 2RNH_2^{(g)} + H_2O$ |

*\* By orange color redox reactions are highlighted, with blue color – non-redox reactions.*

In the case of strongly oxidizing atmosphere, the listed above light-induced processes become oxygen-mediated (Table 2). The reducing agents for $O_2$ could be iodide-anions (2.1) and photoexcited electrons (2.2). In the second case formation of superoxide is expected due to the lower redox potential of $O_2/O_2^-$ with respect to conduction band minimum (CBM) of MAPI.[35] For 2D LHPs the difference between CBM and $O_2/O_2^-$ potential should be even larger, explaining the experimentally observed increase of oxygen-induced degradation rate of layered perovskites. Notably, Aristidou et al. reported that iodide vacancies are the most energetically favored perovskite lattice sites for superoxide formation.[36] In view of this, we propose the following mechanism of oxygen reduction via superoxide formation:

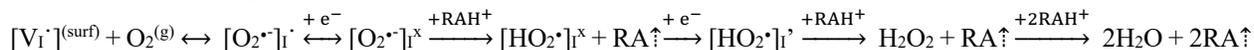

$$[V_I^\bullet]^{(surf)} + O_2^{(g)} \leftrightarrow [O_2^{\bullet-}]_I^\bullet \xrightarrow{+e^-} [O_2^{\bullet-}]_I^x \xrightarrow{+RAH^+} [HO_2^\bullet]_I^x + RA\uparrow \xrightarrow{+e^-} [HO_2^\bullet]_I^{'} \xrightarrow{+RAH^+} H_2O_2 + RA\uparrow \xrightarrow{+2RAH^+} 2H_2O + 2RA\uparrow$$

Highly active superoxide forms interact with vulnerable organic cations deprotonating them and forming $H_2O$ molecules, strongly shifting the reaction "equilibrium" to the right and promoting further water-induced degradation. We believe that reaction (2.2) is more possible than (2.1) due to the fact that $PbI_2$ is the main solid state degradation product instead of lead oxide. Therefore, (2.2), (2.3), and (2.5) reactions are considered to be the main driving force during photooxidation of hybrid lead halide perovskites. Notably, the decrease of oxygen concentration in atmosphere results in the combination of both types of reactions as we can see both $Pb^0$ and $PbI_2$ XRD peaks in perovskite samples after light-soaking under Ar atmosphere with ~ 100 ppm $O_2$ (Figure S13).

CONCLUSIONS

Summing up, we provided a comprehensive stability study of the family of $BA_2MA_{n-1}Pb_nI_{3n+1}$ layered perovskites with n=1-3 and 3D $MAPbI_3$ perovskites. We suggest that in-situ or periodic tracking of perovskite functional properties such as photoluminescence during light or heat treatment is one of the most suitable strategies for comparative stability studies of both 3D and 2D hybrid perovskites with the control of surrounding atmosphere as an essential step preceding these experiments. We found out that the presence of molecular oxygen in the atmosphere during light-soaking tests plays a critical role affecting both photodegradation mechanism and rate of hybrid perovskites and PSCs, as especially important for 2D systems. We proved that the presence of >10 ppm $O_2$ during 2D-PSCs preparation and/or resource testing causes a dramatic loss of their performance. In the inert atmosphere (<10 ppm $O_2$) the layered perovskites become more stable to visible light irradiation and demonstrate a nonmonotonic photostability dependence on the *n* number as follows: $BA_2MA_2Pb_3I_{10}$ > $BA_2MAPb_2I_7$ ~ $MAPbI_3$ > $BA_2PbI_4$. This stability trend could be explained by two antibate factors: (1) the intensification of internal $I^-/I^0$ and $Pb^{2+}/Pb^0$ redox processes along with perovskite bandgap value; (2) the higher volatility of $MA^+$ in MA-abundant compositions (e.g., $MAPbI_3$). As a result, the family of $BA_2MA_{n-1}Pb_nI_{3n+1}$ layered perovskites demonstrates an "island of photostability" for phases with n ≥ 3 in the inert atmosphere and this clue would be practically used in the engineering of solar cell devices.

## ASSOCIATED CONTENT

**Supporting Information**. Supporting Information includes: Experimental Section; comparative table with stability studies on 2D perovskites and PSCs; characterization of prepared samples by XRD, PL and absorption spectroscopy, and EDX; absorption spectra of decane and dependence of the amount of released $I_2$ during perovskites light-soaking; SEM images of perovskite films before and after degradation; XRD of n1 before and after light-soaking in decane; Tauc plot of n2 film during photodegradation in decane and in air; photographs of perovskite films before and after light-soaking in air and in inert glove box; stability results of MAPI and n3 based PSCs in air; photographs and XRD patterns of perovskite samples before and after light-soaking in 99,9% Ar; stability results of MAPI and n3 based PSCs in inert atmosphere; PL spectra of n1, n2, n3, and MAPI samples at different stages of thermal aging and light- soaking tests; the results of two sets of thermal and light stability tests; XRD of perovskite films before and after thermal aging and light-soaking in inert glove box.

## AUTHOR INFORMATION

**Corresponding Author**

* E-Mail: alexey.bor.tarasov@yandex.ru

**Author Contributions**

The manuscript was written through contributions of all the authors. All authors have given approval to the final version of the manuscript.

ACKNOWLEDGMENT

N.N.U., S.A.F., E.M.N., E.A.G., and A.B.T. acknowledge the financial support by the Russian Science Foundation (Project № 19-73-30022). G.G. and A.Z. acknowledge the support from the European Research Council under the European Union's Horizon 2020 research and innovation programme (HYNANO - Grant agreement № 802862). XRD and SEM studies were performed using the equipment of the Joint Research Centre for Physical Methods of Research of Kurnakov Institute of General and Inorganic Chemistry of the Russian Academy of Sciences (JRC PMR IGIC RAS).## REFERENCES